\documentclass[aps,pra,showpacs,onecolumn]{revtex4}
\usepackage{graphicx}

\begin{document}

\title[Comparison of the 1- and 2-order resonances in nonstationary circuit
QED]{Analytical comparison of the first- and second-order resonances for
implementation of the dynamical Casimir effect in nonstationary circuit QED}
\author{E L S Silva$^1$ and A V Dodonov$^{1,2}$}
\affiliation{$^1$ Institute of Physics,
University of Brasilia, 70910-900, Brasilia, Federal District, Brazil}

\affiliation{$^2$ International Center for Condensed Matter Physics,
University of Brasilia, 70910-900, Brasilia, Federal District, Brazil}

\begin{abstract}
We investigate analytically and numerically the nonstationary circuit QED
setup in which $N$ independent qubits interact with a single mode of the
Electromagnetic field confined in a resonator. We consider the harmonic time
modulation of some parameter (atomic transition frequency or the
atom--field coupling strength) and derive the unitary dynamics up to the second
order in the modulation depth for $N=1$ and $N\gg 1$. It is shown that all
the resonant phenomena that occur for modulation frequencies $\sim 2\omega
_{0}$ (where $\omega _{0}$ is the cavity frequency) also occur for the
halved frequencies. However, in the latter case the associated transition
rates are significantly smaller and the modulation of the coupling strength
is less effective. The transition rates are evaluated explicitly and the
prospects of employing the second-order resonances in the phenomena related
to the dynamical Casimir effect are examined.
\end{abstract}


\maketitle
\section{Introduction}

One of the direct proofs of the zero-point fluctuations predicted by quantum
physics is the \emph{dynamical Casimir effect} (DCE) -- a rather broad term
that denotes the creation of quanta from the vacuum state of some field due
to fast changes of the geometry or internal properties of macroscopic or
mesoscopic objects (see \cite{rev1,rev2,rev3,rev4} for reviews). Although
originally ascribed to the generation of photons due to the accelerated
motion of a single mirror \cite{mir1,mir2,mir3} or some cavity wall (where
the photons can be accumulated \cite{cav1,cav2,cav3,cav4}), this phenomenon
has been extended to a variety of other systems, for instance, the phonon
analogues in a Bose-Einstein condensate \cite{BEC1,BEC2}, quantum fluid of
light \cite{QFL} or atomic gas with time-dependent effective charge \cite{ET}%
. Recently two different groups implemented DCE analogues in the circuit
Quantum Electrodynamics (QED) architecture \cite{cir1,cir2,cir4,cir5,cir3},
where the periodic motion of the boundary was simulated by driving magnetic
flux through the SQUIDS (superconducting quantum interference devices)
located at the end \cite{nori-n} or within the coplanar waveguide \cite{meta}%
.

It has been predicted \cite{jpcs,liberato9,JPA,igor,diego,palermo} that the
circuit QED architecture also allows for the implementation of DCE analogue
using a single \emph{artificial atom} -- a macroscopic superconducting
circuit composed of several Josephson junctions that has discrete energy
levels, exhibits coherent quantum oscillations between these levels and can
strongly couple to the electromagnetic field \cite%
{cir3,atoms0,atoms1,atoms2,atoms4}. The internal properties of the
artificial atom can be controlled \emph{in situ} by applying external
electric and magnetic fields \cite{cir2,cir3,atoms3,v1,v2,v3}. Hence, the
generation of photons from vacuum via the modulation of parameters of a
single two-level atom (\emph{qubit}), also initiated in the ground state,
would prove that DCE is an intrinsic effect of the light--matter
interaction, originating from the counter-rotating terms in the interaction
Hamiltonian. Indeed, within a toy model for DCE inside a cavity \cite{igor}, the
modulation of the material properties of a mesoscopic dielectric slab can be
modeled by time-dependent transition frequencies of $N$ atoms, while the
slab's oscillation can be viewed as varying atom--field coupling strengths.

The idea of using qubits with rapidly varying parameters goes beyond the
simple verification of photon generation from vacuum due to the
counter-rotating terms. By carefully adjusting the form of modulation, the
intrinsic nonlinearities in the light--matter interaction \cite{kerr,igor}
may give rise to novel applications in quantum-information processing,
quantum simulations and engineering of nonclassical states of light and
matter \cite{relativistic,processing,entangles,arrays,daniel}. Along with
creation of photon pairs, one can engineer effective interactions that lead
to simultaneous generation of photons and atomic excitations \cite{jpcs} or
coherent annihilation of the system quanta from a known initial state \cite%
{igor,diego}. An intuitive explanation for such behaviors is the
modulation-induced selective coupling between the bare atom--field dressed
states \cite{JPA}. The photon generation and annihilation rates can be
enhanced by using a set of $N$ identical noninteracting qubits, when novel
interaction regimes can be induced as a consequence of a richer spectrum of
the composite system \cite{nlevel,camop,igor,lucas}.

Although the qubit parameters can be readily tuned in the circuit QED setup
\cite{cir2,cir3,atoms3,v1,v2,v3}, one is still left with the technical
difficulty of imposing a periodic modulation with frequency of the order $%
\eta \sim 2\omega _{0}$, where $\omega _{0}/2\pi \sim 10\,$GHz is the cavity
frequency. So the goal
of this paper is to discover whether one could use the halved modulation
frequencies $\eta \sim \omega _{0}$ to achieve similar phenomena via the
second-order resonances. It will be shown analytically in section \ref{desc}
that the answer is yes, and the modulation of the atomic transition
frequency is more efficient than the modulation of the coupling strength to
implement the second-order resonances, whereas for the first-order
resonances they are of the same order \cite{igor,diego}. Moreover, in
section \ref{disc} we shall perform numerical simulations in the presence of
dissipation to assess whether the second-order resonances are feasible with
the current parameters \cite{kerr,art2,art3}. As demonstrated in \ref{apa}, analogous conclusions can be drawn for the case of $N\gg 1$
identical noninteracting qubits with time-dependent parameters. To find the
main conclusions of the work the reader can skip directly to section \ref%
{conc}.

\section{Analytical description of the unitary dynamics}

\label{desc}

We consider a single mode of the electromagnetic field confined in a
microwave waveguide resonator that interacts with $N$ identical qubits via
the dipole interaction. Denoting the cavity frequency, the atomic transition
frequency and the atom--field coupling parameter by $\omega _{0}$, $\Omega $
and $g$, respectively, the dynamics is governed by the generalized Rabi
Hamiltonian \cite{rabi1,rabi2,rabi3} (we set $\hbar =1$)%
\begin{equation}
\hat{H}_{0}=\omega _{0}\hat{n}+i\chi _{0}(\hat{a}^{\dagger 2}-\hat{a}%
^{2})+\sum_{l=1}^{N}[(\Omega /2)\hat{\sigma}_{l,z}+g(\hat{a}+\hat{a}%
^{\dagger })(\hat{\sigma}_{l,+}+\hat{\sigma}_{l,-})]~.  \label{H0}
\end{equation}%
Here $\hat{a}$ and $\hat{a}^{\dagger }$ are the bosonic ladder operators of
the field and $\hat{n}=\hat{a}^{\dagger }\hat{a}$ is the photon number
operator. The qubit operators are $\hat{\sigma}_{l,-}=|g_{l}\rangle \langle
e_{l}|$, $\hat{\sigma}_{l,+}=|e_{l}\rangle \langle g_{l}|$ and $\hat{\sigma}%
_{l,z}=|e_{l}\rangle \langle e_{l}|-|g_{l}\rangle \langle g_{l}|$, where $%
|g_{l}\rangle $ and $|e_{l}\rangle $ denote the\ ground and excited states
of the $l$-th qubit, respectively. The constant squeezing coefficient $\chi
_{0}$ is due to the terms proportional to the square of the vector
potential; it appears naturally when one uses the minimal-coupling
Hamiltonian and the dipole approximation of the first order or higher \cite%
{schleich,igor,man}. We keep this term for the sake of completeness,
although it does not produce any new effect and its contribution is usually
small. We assume that either $\Omega $ or $g$ is modulated externally as $%
X=X_{0}+\varepsilon _{X}\sin (\eta t)$, where $X=\{\Omega ,g\}$, $%
\varepsilon _{X}$ is the corresponding {\em modulation depth} and $\eta $ is the
angular frequency of modulation. The general description of the system
dynamics under the first-order resonances was obtained in \cite{JPA,igor}
(valid even for the simultaneous multi-tone modulation of all the system parameters), so here we concentrate on the second-order effects
whose mathematical analysis is more tricky.

For $N=1$ we work in the Schr\"{o}dinger picture and expand the wavefunction
corresponding to $\hat{H}_{0}$ as \cite{JPA}
\begin{equation}
|\psi (t)\rangle =e^{-it\lambda _{0}}A_{0}(t)|\varphi _{0}\rangle
+\sum_{n=1}^{\infty }\sum_{\mathcal{S}=\pm }e^{-it\lambda _{n,\mathcal{S}%
}}A_{n,\mathcal{S}}(t)|\varphi _{n,\mathcal{S}}\rangle ~,
\end{equation}%
where $\lambda _{n,\mathcal{S}}$ and $|\varphi _{n,\mathcal{S}}\rangle $ are
the $n$-excitations eigenvalues and eigenstates of the bare Jaynes-Cummings
(JC) Hamiltonian \cite{schleich}
\begin{equation}
\hat{H}_{JC}=\omega _{0}\hat{n}+\Omega _{0}|e\rangle \langle e|+g_{0}(\hat{a}%
\hat{\sigma}_{+}+\hat{a}^{\dagger }\hat{\sigma}_{-})~.
\end{equation}%
The index $\mathcal{S}$ labels the different eigenstates with the same
number of excitations $n$.

The well known eigenfrequencies are%
\begin{equation}
\lambda _{0}=0~,~\lambda _{n>0,\mathcal{S}}=\omega _{0}n+\frac{\mathcal{S}%
\beta _{n}-\Delta _{-}}{2}~,~\beta _{n}=\sqrt{\Delta _{-}^{2}+4g_{0}^{2}n}~,~%
\mathcal{S}=\pm ~,  \label{betan}
\end{equation}%
where $\Delta _{-}=\omega _{0}-\Omega _{0}$ is the average detuning. The JC
eigenstates, also known as \emph{dressed states}, are%
\begin{equation}
|\varphi _{0}\rangle =|g,0\rangle ~,~|\varphi _{n>0,\mathcal{S}}\rangle ={\rm s}%
_{n,\mathcal{S}}|g,n\rangle +{\rm c}_{n,\mathcal{S}}|e,n-1\rangle ~,
\label{dress}
\end{equation}%
where we introduced the notation%
\begin{equation}
{\rm s}_{n,+}=\sin \theta _{n},~{\rm s}_{n,-}=\cos \theta _{n},~{\rm c}_{n,+}=\cos
\theta _{n},~{\rm c}_{n,-}=-\sin \theta _{n}~
\end{equation}%
\begin{equation}
\theta _{n>0}=\arctan \frac{\Delta _{-}+\beta _{n}}{2g_{0}\sqrt{n}}~.
\end{equation}

We introduce effective time-dependent probability amplitudes $b(t)$ via the
ansatz%
\begin{equation}
A_{0}(t)=\exp (-it\nu _{0})b_{0}(t)  \label{testa1}
\end{equation}%
\begin{equation}
A_{m>0,\mathcal{T}}(t)=\exp \left( i\Pi _{m,\mathcal{T},\mathcal{T}}^{(k)}%
\frac{\cos \eta t-1}{\eta }\right) \left[ e^{-it\nu _{m,\mathcal{T}}}b_{m,%
\mathcal{T}}(t)-\zeta _{m,\mathcal{T}}^{(k)}(t)e^{-it\nu _{m,-\mathcal{T}%
}}b_{m,-\mathcal{T}}(t)\right] \,,  \label{testa2}
\end{equation}%
where $k=\{\Omega ,g\}$. We defined the small time-dependent coefficient%
\begin{eqnarray}
\zeta _{m,\mathcal{T}}^{(k)}(t) &=&\frac{i\Pi _{m,\mathcal{T},\mathcal{-T}%
}^{(k)}}{2}\exp \left( i\frac{\Pi _{m,\mathcal{T},\mathcal{T}}^{(k)}-\Pi _{m,%
\mathcal{-T},\mathcal{-T}}^{(k)}}{\eta }\right)  \nonumber \\
 &&\times \left[ \frac{e^{it\left( \mathcal{T}\beta _{m}+\eta \right) }-1%
}{\eta +\mathcal{T}\beta _{m}}+\frac{e^{it\left( \mathcal{T}\beta _{m}-\eta
\right) }-1}{\eta -\mathcal{T}\beta _{m}}\right. \\
 &&+\left. \frac{\Pi _{m,\mathcal{T},\mathcal{T}}^{(k)}-\Pi _{m,\mathcal{%
-T},\mathcal{-T}}^{(k)}}{2i\eta }\left( \frac{e^{it\left( \mathcal{T}\beta
_{m}+2\eta \right) }-1}{2\eta +\mathcal{T}\beta _{m}}+\frac{e^{it\left(
\mathcal{T}\beta _{m}-2\eta \right) }-1}{2\eta -\mathcal{T}\beta _{m}}%
\right) \right]  \nonumber
\end{eqnarray}%
and the time-independent matrix elements
\begin{equation}
\Pi _{m,\mathcal{T},\mathcal{S}}^{(k=\Omega )}\equiv \varepsilon _{\Omega
}\langle \varphi _{m,\mathcal{T}}|e\rangle \langle e|\varphi _{m,\mathcal{S}%
}\rangle
\end{equation}%
\begin{equation}
\Pi _{m,\mathcal{T},\mathcal{S}}^{(k=g)}\equiv \varepsilon _{g}\langle
\varphi _{m,\mathcal{T}}|(\hat{a}\hat{\sigma}_{+}+\hat{a}^{\dagger }\hat{%
\sigma}_{-})|\varphi _{m,\mathcal{S}}\rangle ~.
\end{equation}%
These quantities can be calculated in a straightforward manner using the
dressed states.

In equations (\ref{testa1}) and (\ref{testa2}) we included small \lq
intrinsic frequency shifts\rq\ $\nu $ -- they appear due to the elimination
of the rapidly oscillating terms and read \cite{igor}%
\begin{equation}
\nu _{0}=-\sum_{\mathcal{S}=\pm }\frac{{\rm c}_{2,\mathcal{S}}^{2}g_{0}^{2}+2%
{\rm s}_{2,\mathcal{S}}^{2}\chi _{0}^{2}}{\lambda _{2,\mathcal{S}}}~
\end{equation}%
\begin{equation}
\nu _{m>0,\mathcal{T}}=-g_{0}^{2}\sum_{\mathcal{S}=\pm }\left[ \frac{%
\left\vert \mathbf{\bar{\Lambda}}_{m+2,\mathcal{T},\mathcal{S}}\right\vert
^{2}}{\lambda _{m+2,\mathcal{S}}-\lambda _{m,\mathcal{T}}}-\frac{\left\vert
\mathbf{\bar{\Lambda}}_{m,\mathcal{S},\mathcal{T}}\right\vert ^{2}}{\lambda
_{m,\mathcal{T}}-\lambda _{m-2,\mathcal{S}}}\right] ~,  \label{natal2}
\end{equation}%
where we defined%
\begin{equation}
\mathbf{\bar{\Lambda}}_{m+2,\mathcal{T},\mathcal{S}}\equiv \Lambda _{m+2,%
\mathcal{T},\mathcal{S}}-i\frac{\chi _{0}}{g_{0}}L_{m+2,\mathcal{T},\mathcal{%
S}}
\end{equation}%
\begin{equation}
\Lambda _{m+2,\mathcal{T},\mathcal{S}}\equiv \langle \varphi _{m,\mathcal{T}%
}|\hat{a}\hat{\sigma}_{-}|\varphi _{m+2,\mathcal{S}}\rangle ~,~\quad L_{m+2,%
\mathcal{T},\mathcal{S}}\equiv \langle \varphi _{m,\mathcal{T}}|\hat{a}%
^{2}|\varphi _{m+2,\mathcal{S}}\rangle ~.
\end{equation}

After long analytical manipulations in which the off-resonant terms were
consistently eliminated through the rotating-wave approximation (RWA) \cite{JPA},
we find that the effective probability amplitudes obey the differential
equation (to compact the notation we denote $b_{0,\mathcal{T}}\equiv b_{0}$,
$|\varphi _{0,\mathcal{T}}\rangle \equiv |\varphi _{0}\rangle $ and $\lambda
_{0,\mathcal{T}}\equiv \lambda _{0}$)%
\begin{eqnarray}
\dot{b}_{m,\mathcal{T}} &=&\sum_{\mathcal{S}}\left[ \left( \Theta _{m+2,%
\mathcal{T},\mathcal{S}}^{(k)}e^{it\eta }+\Phi _{m+2,{\mathcal{T}},{\mathcal{%
S}}}^{(k)}e^{it2\eta }\right) e^{-it\left( \bar{\lambda}_{m+2,\mathcal{S}}-%
\bar{\lambda}_{m,\mathcal{T}}\right) }b_{m+2,\mathcal{S}}\right.  \nonumber \\
 &&\left. -\left( \Theta _{m,\mathcal{S},\mathcal{T}}^{(k)}e^{it\eta
}+\Phi _{m,{\mathcal{S}},{\mathcal{T}}}^{(k)}e^{it2\eta }\right) ^{\ast
}e^{it\left( \bar{\lambda}_{m,\mathcal{T}}-\bar{\lambda}_{m-2,\mathcal{S}%
}\right) }b_{m-2,\mathcal{S}}\right] ~.  \label{b}
\end{eqnarray}%
Hence both the first- and second-order resonances induce the
transitions between the dressed states differing by two excitations. The
time-independent transition rates corresponding to the first- and
second-order resonances read%
\begin{equation}
\Theta _{m,\mathcal{T},\mathcal{S}}^{(k)}=\frac{g_{0}}{2}\left[ \sum_{%
\mathcal{R}=\pm }\left( \frac{\mathbf{\bar{\Lambda}}_{m,\mathcal{T},\mathcal{%
R}}\Pi _{m,\mathcal{R},\mathcal{S}}^{(k)}}{\eta -\mathcal{S}\delta _{%
\mathcal{R},-\mathcal{S}}\beta _{m}}-\frac{\mathbf{\bar{\Lambda}}_{m,%
\mathcal{R},\mathcal{S}}\Pi _{m-2,\mathcal{T},\mathcal{R}}^{(k)}}{\eta +%
\mathcal{T}\delta _{\mathcal{R},-\mathcal{T}}\beta _{m-2}}\right) -\frac{%
\delta _{k,g}\varepsilon _{g}}{g_{0}}\Lambda _{m,\mathcal{T},\mathcal{S}}%
\right]   \label{c1}
\end{equation}%
\begin{eqnarray}
&\Phi _{m,{\mathcal{T}},{\mathcal{S}}}^{(k)}=\frac{ig_{0}}{4}\left[ \frac{%
\mathbf{\bar{\Lambda}}_{m,\mathcal{T},\mathcal{S}}(\Pi _{m,\mathcal{S},%
\mathcal{S}}^{(k)}-\Pi _{m-2,\mathcal{T},\mathcal{T}}^{(k)})^{2}}{2\eta ^{2}}%
-\frac{\mathbf{\bar{\Lambda}}_{m,-\mathcal{T},-\mathcal{S}}\Pi _{m,-\mathcal{%
S},\mathcal{S}}^{(k)}\Pi _{m-2,\mathcal{T},-\mathcal{T}}^{(k)}}{(\eta -%
\mathcal{S}\beta _{m})(\eta +\mathcal{T}\beta _{m-2})}\right. &  \nonumber \\
&+\frac{\mathbf{\bar{\Lambda}}_{m,\mathcal{T},-\mathcal{S}}\Pi _{m,-%
\mathcal{S},\mathcal{S}}^{(k)}}{\eta }\left( \frac{\Pi _{m,-\mathcal{S},-%
\mathcal{S}}^{(k)}-\Pi _{m-2,\mathcal{T},\mathcal{T}}^{(k)}}{\eta -\mathcal{S%
}\beta _{m}}+\frac{\Pi _{m,\mathcal{S},\mathcal{S}}^{(k)}-\Pi _{m,-\mathcal{S%
},-\mathcal{S}}^{(k)}}{2\eta -\mathcal{S}\beta _{m}}\right) &  \nonumber \\
&-\frac{\mathbf{\bar{\Lambda}}_{m,-\mathcal{T},\mathcal{S}}\Pi _{m-2,%
\mathcal{T},-\mathcal{T}}^{(k)}}{\eta }\left( \frac{\Pi _{m,\mathcal{S},%
\mathcal{S}}^{(k)}-\Pi _{m-2,-\mathcal{T},-\mathcal{T}}^{(k)}}{\eta +%
\mathcal{T}\beta _{m-2}}+\frac{\Pi _{m-2,-\mathcal{T},-\mathcal{T}%
}^{(k)}-\Pi _{m-2,\mathcal{T},\mathcal{T}}^{(k)}}{2\eta +\mathcal{T}\beta
_{m-2}}\right) &  \label{c2} \\
&\left. -\frac{\delta _{k,g}\varepsilon _{g}}{g_{0}}\left( \frac{\Lambda
_{m,\mathcal{T},\mathcal{S}}(\Pi _{m,\mathcal{S},\mathcal{S}}^{(g)}-\Pi
_{m-2,\mathcal{T},\mathcal{T}}^{(g)})}{\eta }+\frac{\Lambda _{m,\mathcal{T},-%
\mathcal{S}}\Pi _{m,-\mathcal{S},\mathcal{S}}^{(g)}}{\eta -\mathcal{S}\beta
_{m}}-\frac{\Lambda _{m,-\mathcal{T},\mathcal{S}}\Pi _{m-2,\mathcal{T},-%
\mathcal{T}}^{(g)}}{\eta +\mathcal{T}\beta _{m-2}}\right) \right] .&  \nonumber
\end{eqnarray}%
For the validity of equation (\ref{b}) the following inequalities are
required%
\begin{eqnarray}
&|\Pi _{m,\mathcal{S},\mathcal{S}}^{(k)}-\Pi _{m,\mathcal{-S},\mathcal{-S}%
}^{(k)}|,|\Pi _{m,\mathcal{S},-\mathcal{S}}^{(k)}|,|\Pi _{m\pm 2,\mathcal{S},%
\mathcal{S}}^{(k)}-\Pi _{m,\mathcal{-S},\mathcal{-S}}^{(k)}|,&  \label{ap1}
\\
&|\Pi _{m\pm 1,\mathcal{S},\mathcal{S}}^{(k)}-\Pi _{m,\mathcal{-S},\mathcal{%
-S}}^{(k)}|,|g_{0}\Lambda _{m+2,\mathcal{T},\mathcal{S}}|,|\chi _{0}L_{m+2,%
\mathcal{T},\mathcal{S}}|\ll \omega _{0}&\,.  \nonumber
\end{eqnarray}

Notice that in equation (\ref{b}) the resonant modulation frequencies
correspond to the difference between two `corrected'\ eigenfrequencies
defined as%
\begin{equation}
\bar{\lambda}_{m,\mathcal{T}}\equiv \lambda _{m,\mathcal{T}}+\nu _{m,%
\mathcal{T}}+\Delta \nu _{m,\mathcal{T}}  \label{lamb}
\end{equation}%
(we denote $\nu _{0,\mathcal{T}}\equiv \nu _{0}$ and $\bar{\lambda}_{0,%
\mathcal{T}}\equiv \bar{\lambda}_{0}$). So the Jaynes-Cummings
eigenfrequencies are corrected by the terms $\nu _{m,\mathcal{T}}$ that
include the standard Bloch-Siegert shift \cite{blais} and a small shift due
to the squeezing coefficient. Within our approach we neglect the additional
frequency shifts $\Delta \nu _{m,\mathcal{T}}$ of the order%
\begin{equation}
O(\Delta \nu _{m,\mathcal{T}})\sim \frac{(\Pi _{m,\mathcal{S},\mathcal{-S}%
}^{(k)})^{2}}{\omega _{0}},\frac{(\Pi _{m\pm 2,\mathcal{S},\mathcal{S}%
}^{(k)}-\Pi _{m,\mathcal{-S},\mathcal{-S}}^{(k)})^{2}}{\omega _{0}}%
~,~k=\{\Omega ,g\}  \label{ap3}
\end{equation}%
and other shifts much smaller than $\nu _{m,\mathcal{T}}$. These terms were
called \lq systematic-error frequency shifts\lq\ (SEFS) in \cite{igor}, since
they appear due to the systematic simplification of the differential
equations for $b_{m,\mathcal{T}}$ using the RWA \cite{JPA}. The knowledge of SEFS is important because they
slightly alter the resonant modulation frequencies, so ultimately they ought
to be found numerically or experimentally to tune the exact resonance.

The functional dependence of the transition rates (\ref{c1}) and (\ref{c2})
on the system parameters is most clearly seen in the particular cases of the
resonant and dispersive regimes studied below.

\subsection{Resonant regime}

\label{res}

For $\Delta _{-}=0$ the corrected eigenfrequencies are $\bar{\lambda}%
_{0}=-(\delta _{+}+\delta _{\chi }/2)$ and $\bar{\lambda}_{m>0,\mathcal{S}%
}=\omega _{0}m+\mathcal{S}g_{0}\sqrt{m}-(\delta _{+}+m\delta _{\chi })$,
where $\delta _{\pm }\equiv g_{0}^{2}/\Delta _{\pm }$,$~\delta _{\chi
}\equiv 4\chi _{0}^{2}/\Delta _{+}$, $\Delta _{+}=\omega _{0}+\Omega _{0}$.
The criteria for validity of our method read: $\varepsilon _{\Omega }$, $%
g_{0}\sqrt{m}$, $\varepsilon _{g}\sqrt{m}$, $\chi _{0}m\ll \omega _{0}$ and $%
O(\Delta \nu _{m,\mathcal{T}})\sim \varepsilon _{\Omega }^{2}/\omega _{0}$%
, $m\varepsilon _{g}^{2}/\omega _{0}$, where $m$ is the number of system
excitations.

Using the JC eigenstates $|\varphi _{m>0,\mathcal{S}}\rangle =(|g,m\rangle +%
\mathcal{S}|e,m-1\rangle )/\sqrt{2}$ we obtain the following transition rates%
\begin{equation}
\Theta _{2,\mathcal{T},\mathcal{S}}^{(k=\Omega )}\simeq \mathcal{S}g_{0}%
\sqrt{2}\left( \frac{\varepsilon _{\Omega }}{8\Omega _{0}}\right) ~,\Phi _{2,%
{\mathcal{T}},{\mathcal{S}}}^{(k=\Omega )}\simeq \mathcal{S}g_{0}\sqrt{2}%
\left( \frac{\varepsilon _{\Omega }}{8\Omega _{0}}\right) ^{2}\left( 3i-%
\mathcal{S}\sqrt{2}\frac{\chi _{0}}{g_{0}}\right)
\end{equation}%
\begin{eqnarray}
\Theta _{m+2,\mathcal{T},{\mathcal{S}}}^{(k=\Omega )} &\simeq &{\mathcal{S}}%
g_{0}\sqrt{m+1}\left( \frac{\varepsilon _{\Omega }}{8\Omega _{0}}\right) ~,~
\\
\Phi _{m+2,{\mathcal{T}},{\mathcal{S}}}^{(k=\Omega )} &\simeq &\mathcal{S}%
g_{0}\sqrt{m+1}\left( \frac{\varepsilon _{\Omega }}{8\Omega _{0}}\right) ^{2}%
\left[ 2i-\frac{2\chi _{0}}{g_{0}}(\mathcal{S}\sqrt{m+2}+\mathcal{T}\sqrt{m})%
\right]
\end{eqnarray}%
\begin{equation}
\Theta _{2,\mathcal{T},{\mathcal{S}}}^{(k=g)}\simeq -{\mathcal{S}}\frac{g_{0}%
}{\sqrt{2}}\left( \frac{\varepsilon _{g}}{2g_{0}}\right) ~,~\Phi _{2,{%
\mathcal{T}},{\mathcal{S}}}^{(k=g)}\simeq -{\mathcal{S}}\frac{g_{0}}{\sqrt{2}%
}\left( \frac{\varepsilon _{g}}{2g_{0}}\right) ^{2}\frac{i{\mathcal{S}}\sqrt{%
2}g_{0}}{\omega _{0}}
\end{equation}%
\begin{eqnarray}
\Theta _{m+2,\mathcal{T},{\mathcal{S}}}^{(k=g)} &\simeq &-{\mathcal{S}}\frac{%
g_{0}}{2}\sqrt{m+1}\left( \frac{\varepsilon _{g}}{2g_{0}}\right) ~,~ \\
\Phi _{m+2,{\mathcal{T}},{\mathcal{S}}}^{(k=g)} &\simeq &-{\mathcal{S}}\frac{%
g_{0}}{2}\sqrt{m+1}\left( \frac{\varepsilon _{g}}{2g_{0}}\right) ^{2}\frac{%
ig_{0}}{\omega _{0}}({\mathcal{S}}\sqrt{m+2}-\mathcal{T}\sqrt{m}),
\end{eqnarray}%
where $m>0$. Notice that for $k=\Omega $ the second-order resonances are one
order of magnitude weaker than the first-order resonances, while for $k=g$
the transition rates are even smaller due to the additional factor $g_{0}/\omega
_{0}\ll 1$.

\subsection{Dispersive regime}

\label{dr}

For $|\Delta _{-}|/2\gg g_{0}\sqrt{m}$ we obtain after expanding $\beta _{m}$
in equation (\ref{betan})%
\begin{eqnarray}
\bar{\lambda}_{m,\mathcal{D}} &\simeq &(\omega _{0}+\delta _{-}-\delta
_{+}-\delta _{\chi })m-\alpha m^{2}-\delta _{+}-\frac{1}{2}\delta _{\chi }~
\\
~\bar{\lambda}_{m>0,\mathcal{-D}} &\simeq &(\omega _{0}-\delta _{-}+\delta
_{+}-\delta _{\chi })m-\Delta _{-}+\alpha m^{2}-\delta _{+}+\frac{1}{2}%
\delta _{\chi }\,,  \nonumber
\end{eqnarray}%
where we denote $\bar{\lambda}_{0,\mathcal{D}}\equiv \bar{\lambda}_{0}$. $%
\mathcal{D}=\Delta _{-}/|\Delta _{-}|=\pm $ is the \lq detuning symbol\rq\
and the effective single-photon Kerr nonlinearity strength is $\alpha =g_{0}^{4}/\Delta
_{-}^{3}$ \cite{igor,blais}. To the first order in $g_{0}/\Delta _{-}$ the
JC eigenstates are%
\begin{eqnarray}
|\varphi _{m\geq 0,\mathcal{D}}\rangle &\simeq &\left( |g,m\rangle +%
\frac{g_{0}}{\Delta _{-}}\sqrt{m}|e,m-1\rangle \right) \\
|\varphi _{m>0,\mathcal{-D}}\rangle &\simeq &-\mathcal{D}\left( |e%
,m-1\rangle -\frac{g_{0}}{\Delta _{-}}\sqrt{m}|g,m\rangle \right) 
\nonumber
\end{eqnarray}%
(we denote $|\varphi _{0,\mathcal{D}}\rangle \equiv |\varphi _{0}\rangle $)
and the criteria for validity of our approach are the same as in section \ref%
{res} plus $|\Delta _{-}|\ll \omega _{0}.$ In this regime one can
distinguish three qualitatively different behaviors: the
Anti-Jaynes-Cummings (AJC) behavior, DCE and Anti-DCE.

In the \emph{AJC regime} \cite{jpcs} one couples the states $\{|\varphi _{m,\mathcal{D}%
}\rangle ,|\varphi _{m+2,-\mathcal{D}}\rangle \}$ when the modulation
frequency is roughly $\eta \simeq \Delta _{+}$ (1-order resonance) or $\eta
\simeq \Delta _{+}/2$ (2-order resonance). This corresponds roughly to the
transition $|g,m\rangle \leftrightarrow |e,m+1\rangle $, and the approximate
expressions for the transition rates are (for $m\geq 0$)%
\begin{eqnarray}
\Theta _{m+2,\mathcal{D},\mathcal{-D}}^{(k=\Omega )} &\simeq &-\mathcal{D}%
g_{0}\sqrt{m+1}\left( \frac{\varepsilon _{\Omega }}{2\Delta _{+}}\right) ~~
\label{s1} \\
\Phi _{m+2,\mathcal{D},{-}\mathcal{D}}^{(k=\Omega )} &\simeq &-\mathcal{D}%
g_{0}\sqrt{m+1}\left( \frac{\varepsilon _{\Omega }}{2\Delta _{+}}\right)
^{2}\times 2i  \nonumber
\end{eqnarray}%
\begin{eqnarray}
\Theta _{m+2,\mathcal{D},\mathcal{-D}}^{(k=g)} &\simeq &\mathcal{D}g_{0}%
\sqrt{m+1}\left( \frac{\varepsilon _{g}}{2g_{0}}\right) ~ \\
\Phi _{m+2,\mathcal{D},{-}\mathcal{D}}^{(k=g)} &\simeq &\mathcal{D}g_{0}%
\sqrt{m+1}\left( \frac{\varepsilon _{g}}{2g_{0}}\right) ^{2}\frac{g_{0}}{%
\Delta _{+}}\frac{4g_{0}(m+1)}{i\Delta _{-}}\,.  \nonumber
\end{eqnarray}

In the \emph{DCE behavior} one couples the states $\{|\varphi _{m,\pm
\mathcal{D}}\rangle ,|\varphi _{m+2,\pm \mathcal{D}}\rangle ,|\varphi
_{m+4,\pm \mathcal{D}}\rangle ,\ldots \}$ when the modulation frequency is
close to $\eta \simeq 2\omega _{0}$ or $\eta \simeq \omega _{0}$.
Intuitively, these transitions correspond roughly to $|g,m\rangle
\leftrightarrow |g,m+2\rangle \leftrightarrow |g,m+4\rangle \cdots $ for the
states $(+\mathcal{D})$ or $|e,m\rangle \leftrightarrow |e,m+2\rangle
\leftrightarrow |e,m+4\rangle \cdots $ for the states $(-\mathcal{D})$, so
photon pairs can be generated from the initial vacuum field state (accompanied by low atomic excitation). A
thorough analysis of the system dynamics under the 1-order resonance was
performed in \cite{igor}, where one showed that the average photon number
and the degree of quadrature squeezing undergo saturation effects due to the
effective Kerr nonlinearity, while the dynamics exhibits a collapse-revival
behavior. The transition rates read ($m\geq 0$)

\begin{eqnarray}
\Theta _{m+2,\mathcal{D},\mathcal{D}}^{(\Omega )} &\simeq &\delta _{-}\sqrt{%
(m+1)(m+2)}\left( \frac{\varepsilon _{\Omega }}{2\Delta _{+}}\right) ~ \\
\Phi _{m+2,\mathcal{D},\mathcal{D}}^{(\Omega )} &\simeq &\delta _{-}\sqrt{%
(m+1)(m+2)}\left( \frac{\varepsilon _{\Omega }}{2\Delta _{+}}\right) ^{2}%
\frac{\Delta _{+}}{\Omega _{0}}\left[ i-\frac{2\chi _{0}}{\Delta _{-}}(m+1)%
\right]  \nonumber
\end{eqnarray}%
\begin{eqnarray}
\Theta _{m+2,\mathcal{D},\mathcal{D}}^{(g)} &\simeq &-\delta _{-}\sqrt{%
(m+1)(m+2)}\left( \frac{\varepsilon _{g}}{2g_{0}}\right) \frac{2\Omega _{0}}{%
\Delta _{+}}~ \\
\Phi _{m+2,\mathcal{D},\mathcal{D}}^{(g)} &\simeq &-\delta _{-}\sqrt{%
(m+1)(m+2)}\left( \frac{\varepsilon _{g}}{2g_{0}}\right) ^{2}\frac{i\Delta
_{-}}{\Omega _{0}}  \nonumber
\end{eqnarray}%
\begin{eqnarray}
\Theta _{m+2,\mathcal{-D},\mathcal{-D}}^{(\Omega )} &\simeq &-\delta _{-}%
\sqrt{m(m+1)}\left( \frac{\varepsilon _{\Omega }}{2\Delta _{+}}\right) ~ \\
\Phi _{m+2,{-}\mathcal{D},{-}\mathcal{D}}^{(\Omega )} &\simeq &-\delta _{-}%
\sqrt{m(m+1)}\left( \frac{\varepsilon _{\Omega }}{2\Delta _{+}}\right) ^{2}%
\frac{\Delta _{+}}{\Omega _{0}}\left[ i+\frac{2\chi _{0}}{\Delta _{-}}(m+1)%
\right]  \nonumber
\end{eqnarray}%
\begin{eqnarray}
\Theta _{m+2,\mathcal{-D},\mathcal{-D}}^{(g)} &\simeq &\delta _{-}\sqrt{%
m(m+1)}\left( \frac{\varepsilon _{g}}{2g_{0}}\right) \frac{2\Omega _{0}}{%
\Delta _{+}}~ \\
\Phi _{m+2,{-}\mathcal{D},{-}\mathcal{D}}^{(g)} &\simeq &\delta _{-}\sqrt{%
m(m+1)}\left( \frac{\varepsilon _{g}}{2g_{0}}\right) ^{2}\frac{i\Delta _{-}}{%
\Omega _{0}}~.  \nonumber
\end{eqnarray}

The \emph{Anti-DCE behavior} \cite{igor,diego,lucas} couples the states $%
\{|\varphi _{m+2,\mathcal{D}}\rangle ,|\varphi _{m,-\mathcal{D}}\rangle \}$
and occurs for $\eta \simeq (3\omega _{0}-\Omega _{0})$ or $\eta \simeq
(3\omega _{0}-\Omega _{0})/2$. Since this regime corresponds to the coherent
annihilation of two excitations when the system is initiated in some state $%
(+\mathcal{D})$, roughly represented by the transition $|g,m+2\rangle
\leftrightarrow |e,m-1\rangle $, the nickname Anti-DCE seems appropriate.
The approximate transition rates are%
\begin{eqnarray}
\Theta _{m+2,\mathcal{-D},\mathcal{D}}^{(\Omega )} &\simeq &\frac{1}{2}%
\mathcal{D}\delta _{-}\frac{g_{0}}{\Delta _{-}}\sqrt{m(m+1)(m+2)}\left(
\frac{\varepsilon _{\Omega }}{2\omega _{0}}\right) \\
\Phi _{m+2,{-}\mathcal{D},\mathcal{D}}^{(\Omega )} &\simeq &\frac{1}{2}%
\mathcal{D}\delta _{-}\frac{g_{0}}{\Delta _{-}}\sqrt{m(m+1)(m+2)}\left(
\frac{\varepsilon _{\Omega }}{2\omega _{0}}\right) ^{2}\left( i-\frac{2\chi
_{0}}{\Delta _{-}}\right)  \nonumber
\end{eqnarray}%
\begin{eqnarray}
\Theta _{m+2,\mathcal{-D},\mathcal{D}}^{(g)} &\simeq &-\mathcal{D}\delta _{-}%
\frac{g_{0}}{\Delta _{-}}\sqrt{m(m+1)(m+2)}\left( \frac{\varepsilon _{g}}{%
2g_{0}}\right) \frac{\Omega _{0}}{\omega _{0}}  \label{s2} \\
\Phi _{m+2,{-}\mathcal{D},\mathcal{D}}^{(g)} &\simeq &-\mathcal{D}\delta _{-}%
\frac{g_{0}}{\Delta _{-}}\sqrt{m(m+1)(m+2)}\left( \frac{\varepsilon _{g}}{%
2g_{0}}\right) ^{2}\frac{2i\Delta _{-}}{\omega _{0}}.  \nonumber
\end{eqnarray}%
We notice that Anti-DCE is one order of magnitude weaker than DCE and only
couples one pair of states, so it is harder to implement than all the other
phenomena discussed above \cite{diego}.

From equations (\ref{s1}) -- (\ref{s2}) we see that under the 2-order
resonances the modulation of $\Omega $ is more efficient than the modulation
of $g$, although for the 1-order resonances both modulations give rise to
transition rates with the same order of magnitude [assuming $O(\varepsilon
_{\Omega }/\Omega _{0})\sim O(\varepsilon _{g}/g_{0})$].

\section{Discussion}

\label{disc}

In the preceding section we have evaluated explicitly the transition rates
for single-qubit DCE and related phenomena under the 1- and 2-order
resonances, showing that the modulation of $\Omega $ is more efficacious to
achieve the second-order resonances. The first natural question that arises
is whether such conclusion holds if one increases the number of qubits,
e.g., by using a cloud of cold polar molecules trapped above the resonator
\cite{polar1,polar2}. We treated this issue in the limit $N\gg 1$, when the
Holstein-Primakoff transformation \cite{HP} permits to obtain relatively simple
expressions for effective Hamiltonians. After long calculations summarized
in \ref{apa} we confirmed that $\Omega $-modulation is again more efficient
than $g$-modulation, and the associated transition rates have 
the functional dependence similar to the one found for $N=1$.

The second question is whether the second-order resonances are feasible
with the current or near-future experimental parameters in the single-qubit
circuit QED. Taking into account the realistic parameters $\omega _{0}/2\pi
=10\,$GHz, $g_{0}/\omega _{0}=5\times 10^{-2}$ and $\Delta _{-}=0$ ($\Delta
_{-}=8g_{0}$) in the resonant (dispersive) regime \cite{cir2,cir3,cir5}, we
shall calculate the transition rates for the $\Omega $-modulation assuming
the perturbation depth $\varepsilon _{\Omega }/\Omega _{0}=5\times 10^{-2}$
and setting $\chi _{0}=0$, recalling that for $g$-modulation the
corresponding rates are much smaller. We shall denote the transition rates
under the 1- and 2-order resonances by $\theta _{1}\equiv |\Theta _{M+2,%
\mathcal{T},\mathcal{S}}^{(\Omega )}|$ and $\theta _{2}\equiv |\Phi _{M+2,{%
\mathcal{T}},{\mathcal{S}}}^{(\Omega )}|$, where $M$ denotes the relevant
number of the system excitations.

Since one of the main features of DCE is the photon generation from vacuum,
we consider the initial zero-excitations state (ZES) $|g,0\rangle $. In the
resonant and AJC regimes we obtain $\theta _{1}/g_{0}\approx 9\times 10^{-3}$
and $\theta _{2}/g_{0}\approx 2\times 10^{-4}$, while for the DCE regime $%
\theta _{1}/g_{0}\approx 2\allowbreak \times 10^{-3}$ and $\theta
_{2}/g_{0}\approx 4\allowbreak \times 10^{-5}$. Finally, for Anti-DCE and the
initial state with $m\approx 5$ excitations we have $\theta _{1}/g_{0}\approx
3\times 10^{-4}$ and $\theta _{2}/g_{0}\approx 4\times 10^{-6}$. These values
must be compared with the dominant dissipative parameters -- the cavity
(qubit) damping rate $\kappa $ ($\gamma $) and the qubit's pure dephasing
rate $\gamma _{\phi }$. Considering the state-of-the-art parameters $%
\kappa \sim \gamma \sim \gamma _{\phi }\sim 5\times 10^{-5}g_{0}$ \cite%
{kerr,art2,art3}, it seems that the photon generation from vacuum using the
2-order resonances is possible in systems with weak dissipation.

\begin{figure}[tbh]
\begin{center}
\includegraphics[width=0.9\textwidth]{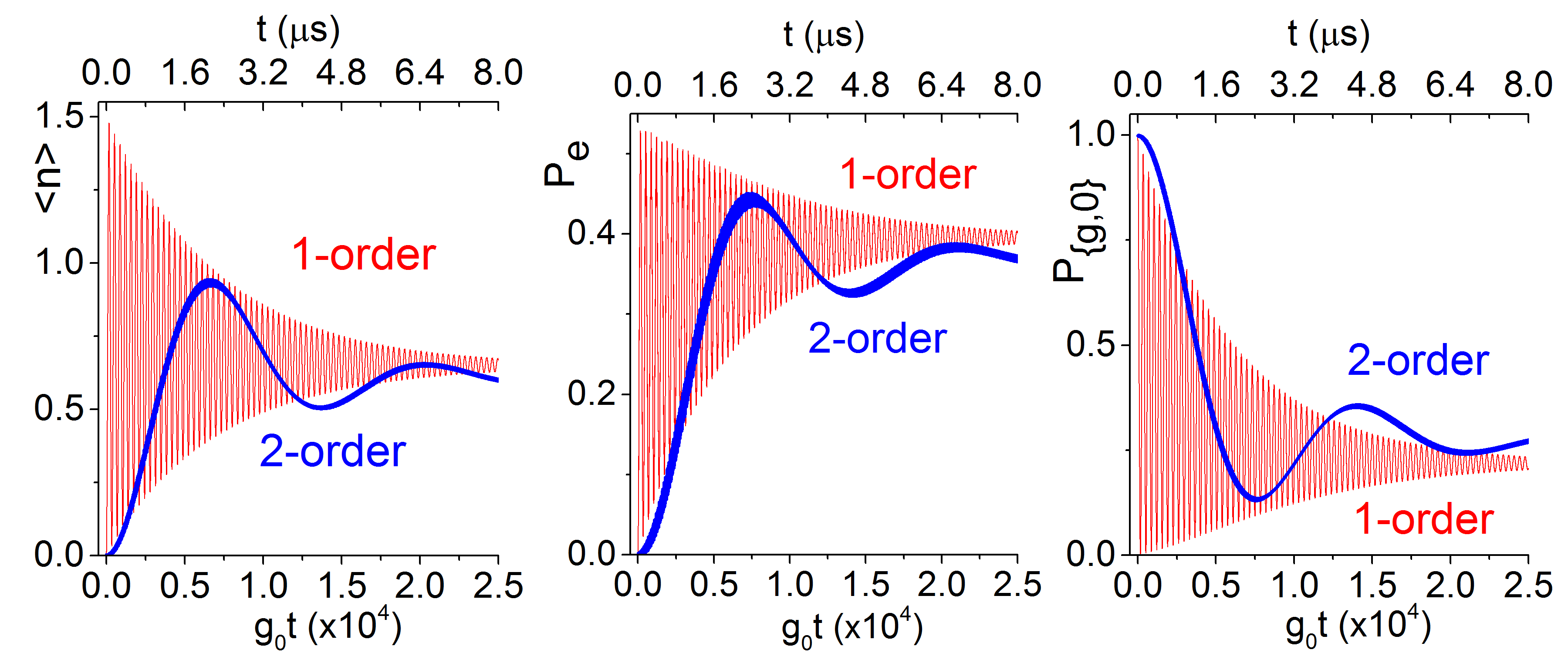} {}
\end{center}
\caption{(Color online) Dynamics of $\langle \hat{n}\rangle $, $P_{e}$ and $%
P_{\{g,0\}}$ for photon generation from the ZES in the resonant regime. The
resonant modulation frequencies found numerically are $\protect\eta =(%
\protect\lambda _{2,+}-\protect\lambda _{0}-7\times 10^{-2}\protect\delta %
_{+})/K$, where $K=1$ or $K=2$.}
\label{fig1}
\end{figure}

\begin{figure}[tbh]
\begin{center}
\includegraphics[width=0.9\textwidth]{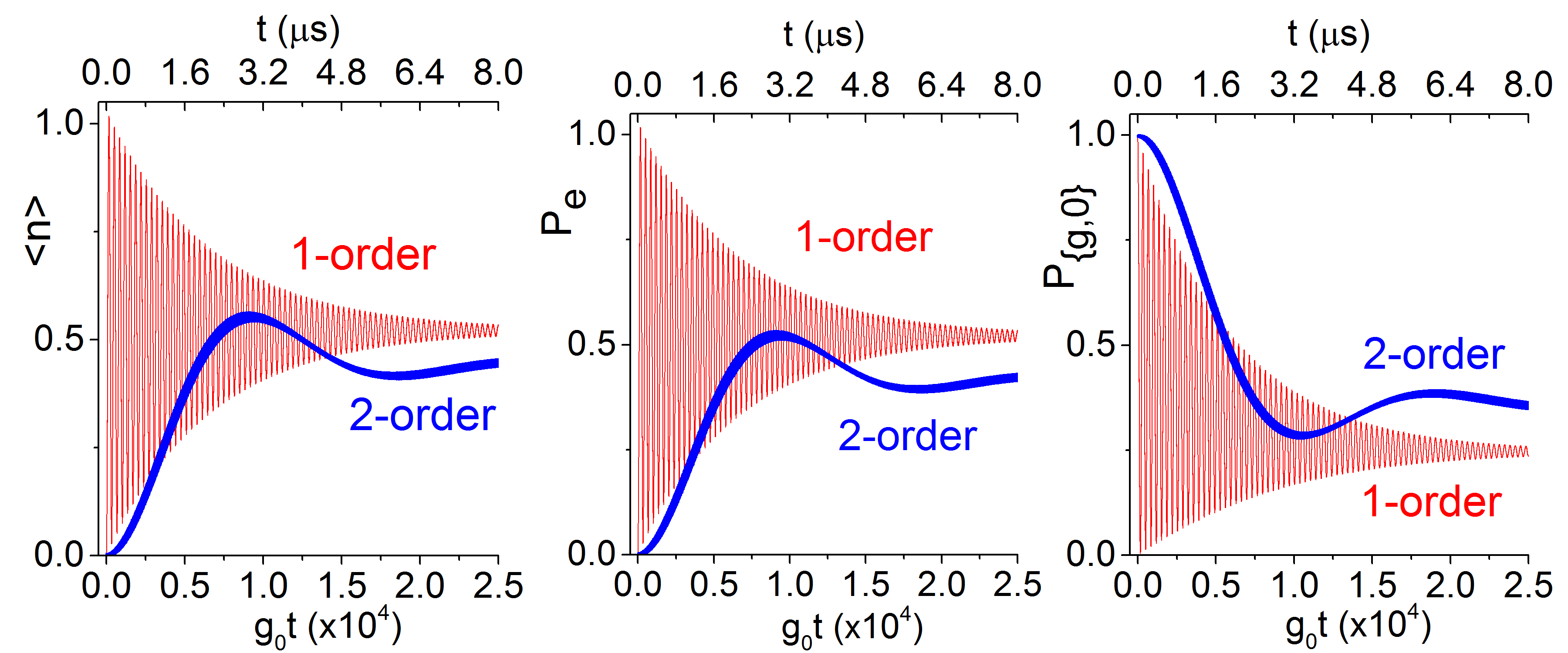} {}
\end{center}
\caption{(Color online) Similar to figure \protect\ref{fig1}, but for the
AJC regime. The resonant modulation frequencies are $%
\protect\eta =(\protect\lambda _{2,-\mathcal{D}}-\protect\lambda _{0}+1.916\,%
\protect\delta _{+})/K$.}
\label{fig2}
\end{figure}

\begin{figure}[tbh]
\begin{center}
\includegraphics[width=0.9\textwidth]{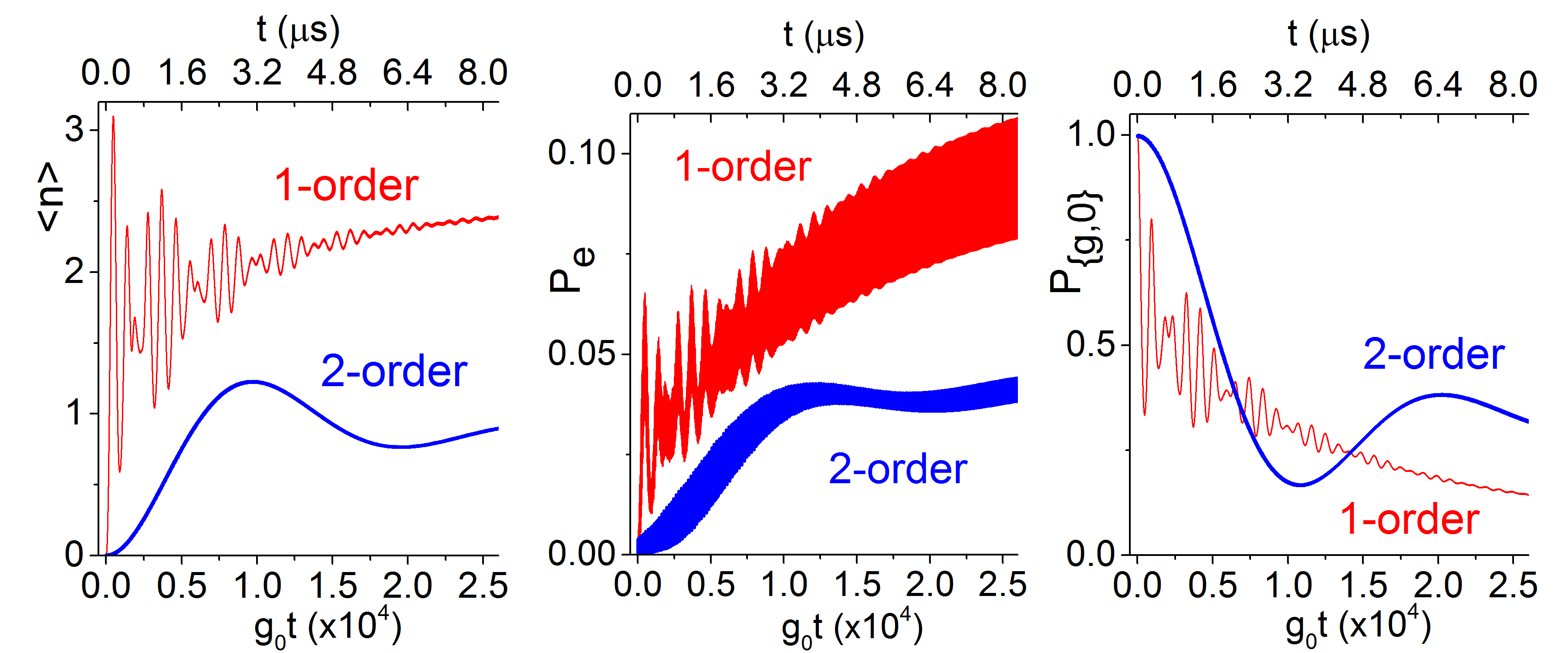} {}
\end{center}
\caption{(Color online) Similar to figure \protect\ref{fig1}, but for the
DCE regime. The resonant modulation frequencies are $\protect\eta =(\protect%
\lambda _{2,\mathcal{D}}-\protect\lambda _{0}-1.93\,\protect\delta _{+})/K$,
$K=1$,$2$. }
\label{fig3}
\end{figure}

To estimate the actual behavior one has to solve the master equation for the
density matrix%
\begin{equation}
d\hat{\rho}/dt=-i[\hat{H},\hat{\rho}]+\hat\mathcal{{L}}\hat{\rho}\,,
\label{mer}
\end{equation}%
where the Liouvillian superoperator $\hat\mathcal{{L}}$ depends on the
system-reservoir interaction. We assume the Markovian regime and employ the
{\em standard master equation} (SME) of Quantum Optics, since recent studies \cite%
{diego,palermo} have shown that for the parameters considered here it
reproduces quite well the results of a more sophisticated microscopic model
\cite{blais}. At zero temperature our dissipative kernel reads \cite{vogel}%
\begin{equation}
\hat\mathcal{{L}}\hat{\rho}=\kappa \emph{D}[\hat{a}]\hat{\rho}+\gamma \emph{D%
}[\hat{\sigma}_{-}]\hat{\rho}+\frac{\gamma _{\phi }}{2}\emph{D}[\hat{\sigma}%
_{z}]\hat{\rho}\,,
\end{equation}%
where $\emph{D}[\hat{O}]\hat{\rho}\equiv \frac{1}{2}(2\hat{O}\hat{\rho}\hat{O%
}^{\dagger }-\hat{O}^{\dagger }\hat{O}\hat{\rho}-\hat{\rho}\hat{O}^{\dagger }%
\hat{O})$ is the Lindbladian superoperator and $\hat{\sigma}_{z}\equiv
|e\rangle \langle e|-|g\rangle \langle g|$. Typical dynamics under the 1-
and 2-order resonances in the resonant, AJC and DCE regimes is displayed in
figures \ref{fig1}, \ref{fig2} and \ref{fig3}, respectively, where equation (%
\ref{mer}) was solved numerically for the initial ZES. The plots illustrate
the evolution of the average photon number $\langle \hat{n}\rangle $, the
atomic excitation probability $P_{e}=\mathrm{Tr}[|e\rangle \langle e|\hat{%
\rho}]$ and the probability of the ZES $P_{\{g,0\}}=\langle g,0|\hat{\rho}%
|g,0\rangle $ for parameters $g_{0}/\omega _{0}=5\times 10^{-2}$, $\kappa
=\gamma =\gamma _{\phi }=10^{-4}g_{0}$, $\Delta _{-}=0$ ($\Delta _{-}=8g_{0}$%
) in the resonant (dispersive) regime, $\varepsilon _{\Omega }/\Omega
_{0}=5\times 10^{-2}$ in figures \ref{fig1} -- \ref{fig2} and $\varepsilon
_{\Omega }/\Omega _{0}=10^{-1}$ in figure \ref{fig3}. One can see that a
measurable amount of cavity and atomic excitations can still be generated
under the 2-order resonances on the timescale of a few microseconds, and the
oscillatory behavior predicted by equation (\ref{b}) is partially preserved.
Moreover, the probability of measuring states other than the initial ZES can
become larger than $70\%$ for some time intervals.

From figures \ref{fig1}--\ref{fig2} we see that, curiously, the asymptotic
values are quite close under the 1- and 2-order resonances. To explain such
behavior, one can derive closed analytical expressions by writing the
differential equations for the density matrix elements in the dressed basis
and performing RWA. Following the steps of \cite{diego}, after long
calculations we find that for the \emph{resonant regime} (section \ref{res})
and $\kappa =\gamma =\gamma _{\phi }$ we obtain for $t\rightarrow \infty $%
\begin{equation}
\left\langle \hat{n}\right\rangle ^{(\infty )}\simeq \frac{15}{23+9(\gamma
/\theta )^{2}}~,~P_{e}^{(\infty )}\simeq \frac{3}{5}\left\langle
n\right\rangle _{\infty }~,~P_{\{g,0\}}^{(\infty )}\simeq \frac{5+9(\gamma
/\theta )^{2}}{23+9(\gamma /\theta )^{2}}~,
\end{equation}%
where $\theta $ stands for $\theta _{1}$ or $\theta _{2}$, depending on the
order of resonance. Another relevant scenario is the case of negligible
cavity damping, $\kappa =0$, when for $\gamma =\gamma _{\phi }$%
\begin{equation}
\left\langle \hat{n}\right\rangle ^{(\infty )}\simeq \frac{6}{8+(3\gamma
/4\theta )^{2}}~,~P_{e}^{(\infty )}\simeq \frac{1}{2}\left\langle
n\right\rangle ^{(\infty )}~,~P_{\{g,0\}}^{(\infty )}\simeq \frac{2+(3\gamma
/4\theta )^{2}}{8+(3\gamma /4\theta )^{2}}\,.
\end{equation}%
Similarly, for the \emph{AJC regime} (section \ref{dr}) we obtain for $%
\kappa =\gamma =\gamma _{\phi }$%
\begin{equation}
\left\langle \hat{n}\right\rangle ^{(\infty )}\simeq P_{e}^{(\infty )}\simeq %
\left[ 2+(\gamma /\theta )^{2}\right] ^{-1}~,~P_{\{g,0\}}^{(\infty )}\simeq
\frac{1}{2}\frac{1+2(\gamma /\theta )^{2}}{2+(\gamma /\theta )^{2}}
\end{equation}%
and for $\kappa =0$, $\gamma =\gamma _{\phi }$%
\begin{equation}
\left\langle \hat{n}\right\rangle ^{(\infty )} \simeq \left[ 1+\left(
\frac{g_{0}}{\Delta _{-}}\right) ^{2}\left( \frac{3\gamma }{2\theta }\right)
^{2}\right] ^{-1}~,~P_{e}^{(\infty )}\simeq 6\left( \frac{g_{0}}{\Delta _{-}}%
\right) ^{2}\left\langle \hat{n}\right\rangle ^{(\infty )}~,~~ 
P_{\{g,0\}}^{(\infty )} \simeq \left( \frac{g_{0}}{\Delta _{-}}\right) ^{2}%
\frac{3+(3\gamma /2\theta )^{2}}{1+\left( g_{0}/\Delta _{-}\right)
^{2}(3\gamma /2\theta )^{2}}\,.
\end{equation}%
These expressions fit surprisingly well the numerical data and explain why
the asymptotic values are so close under the 1- and 2-order resonances:  we
have $(\gamma /\theta _{1})^{2}\approx 10^{-4}$ and $(\gamma /\theta
_{2})^{2}\approx 0.25$, so the contributions of these terms are quite small.

\section{Conclusions}

\label{conc}

In this work we obtained approximate analytical expressions for the dynamics
of the nonstationary circuit QED system under the first- and second-order
resonances. We considered a \emph{weak} harmonic modulation of either the
atomic transition frequency ($\Omega $) or the atom--field coupling strength
($g$) of a single or $N\gg 1$ identical qubits, demonstrating that all the
phenomena that occur for the modulation frequency $\eta \sim 2\omega _{0}$
(1-order resonances) also occur for $\eta \sim \omega _{0}$ (2-order
resonances), where $\omega _{0}$ is the cavity frequency. It turned out that
the modulation of $\Omega $ is more efficient for achieving the second-order
resonances, in contrast to the first-order resonances for which both
modulations produce similar transition rates.

We solved numerically the \lq standard master equation\rq\ of quantum optics to
assess the feasibility of generating excitations from vacuum due to the
counter-rotating terms under the second-order resonances. As demonstrated by
figures \ref{fig1}--\ref{fig3}, for small but realistic values of the
dissipation parameters our proposal is realizable on the timescale of a few
microseconds provided the modulation frequency is accurately tuned (with the
absolute precision $\lesssim 10^{-5}\omega _{0}$). However, the benefit of
lowering the modulation frequency by a factor of two is countered by the
decrease of the transition rates by at least one order of magnitude, whence
the dissipation plays a major role and partially destroys the oscillatory behavior expected from
the unitary dynamics. Nevertheless, our results indicate that it is worth to
pursue the second-order resonances in nonstationary circuit QED systems, as
potential stronger modulation depths and atom--field coupling parameter could make them a viable mechanism of coherent selective
coupling between the system dressed states.

\begin{acknowledgments}
 ELSS acknowledges financial support by CAPES (Brazilian agency). AVD
acknowledges a support of the Brazilian agency CNPq (Conselho Nacional de
Desenvolvimento Cient\'{\i}fico e Tecnol\'{o}gico).
We are grateful to L. C. Monteiro for some analytical evaluations at the
initial stage of the work.
\end{acknowledgments}

\appendix{}

\section{Approximate expressions for $N\gg 1$}

\label{apa}

For $N\gg 1$ identical noninteracting qubits we define the collective atomic
operators via the Holstein--Primakoff transformation \cite{HP}%
\begin{equation}
\sum_{j=1}^{N}\hat{\sigma}_{+}^{(j)}=\hat{b}^{\dagger }\sqrt{N-\hat{b}%
^{\dagger }\hat{b}}~,~\sum_{j=1}^{N}\hat{\sigma}_{-}^{(j)}=\sqrt{N-\hat{b}%
^{\dagger }\hat{b}}\hat{b}~,~\sum_{j=1}^{N}\hat{\sigma}_{z}^{(j)}=2\hat{b}%
^{\dagger }\hat{b}-N,
\end{equation}%
where the ladder operators $\hat{b}$ and $\hat{b}^{\dagger }$ satisfy the
bosonic commutation relation. To the first order in $\hat{b}^{\dagger }\hat{b%
}/N$ the Hamiltonian (\ref{H0}) becomes $\hat{H}=\hat{H}_{JC}+\hat{H}_{G}+%
\hat{H}_{NG}+\hat{H}_{m}$, where%
\begin{eqnarray}
\hat{H}_{JC} &=&\omega _{0}\hat{n}+\Omega _{0}\hat{b}^{\dagger }\hat{b}+%
\tilde{g}_{0}(\hat{a}\hat{b}^{\dagger }+\hat{a}^{\dagger }\hat{b})  \nonumber \\
\hat{H}_{G} &=&\tilde{g}(\hat{a}\hat{b}+\hat{a}^{\dagger }\hat{b}^{\dagger
})+i\chi _{0}(\hat{a}^{\dagger 2}-\hat{a}^{2})  \label{NG} \\
\hat{H}_{NG} &=&-\frac{\tilde{g}}{2N}(\hat{a}+\hat{a}^{\dagger })(\hat{b}%
^{\dagger 2}\hat{b}+\hat{b}^{\dagger }\hat{b}^{2})\,.  \nonumber
\end{eqnarray}%
$\hat{H}_{m}=\varepsilon _{\Omega }\sin (\eta t)\hat{b}^{\dagger }\hat{b}$
for the $\Omega $-modulation and $\hat{H}_{m}=\tilde{\varepsilon}_{g}\sin
(\eta t)(\hat{a}\hat{b}^{\dagger }+\hat{a}^{\dagger }\hat{b})$ for the $g$%
-modulation. We also defined the collective coupling parameter $\tilde{g}%
\equiv \sqrt{N}g\equiv \tilde{g}_{0}+\tilde{\varepsilon}_{g}f_{g}$.

We write the solution in the Heisenberg picture as \cite{JPA,igor}%
\begin{equation}
\hat{a}=\beta ^{-1}e^{-it\Delta _{+}/2}\left[ (\beta _{+}\hat{A}_{h}+\tilde{g%
}_{0}\hat{B}_{h})e^{-it\beta /2}+(\beta _{-}\hat{A}_{h}-\tilde{g}_{0}\hat{B}%
_{h})e^{it\beta /2}\right]  \label{an1}
\end{equation}%
\begin{equation}
\hat{b}=\beta ^{-1}e^{-it\Delta _{+}/2}\left[ (\beta _{-}\hat{B}_{h}+\tilde{g%
}_{0}\hat{A}_{h})e^{-it\beta /2}+(\beta _{+}\hat{B}_{h}-\tilde{g}_{0}\hat{A}%
_{h})e^{it\beta /2}\right] \,,  \label{an3}
\end{equation}%
where $\beta \equiv \sqrt{\Delta _{-}^{2}+4\tilde{g}_{0}^{2}}$, $\beta _{\pm
}=(\beta \pm \Delta _{-})/2$ and the independent operators $\hat{A}_{h}$, $%
\hat{B}_{h}$ also satisfy the bosonic commutation relations. Next we propose
the ansatz (for $k=\Omega ,g$)

\begin{equation}
\hat{A}_{h}=\exp [i\mathcal{F}_{A}^{(k)}]\times \left[ \hat{A}e^{it\left(
\tilde{\delta}_{+}+\delta _{\chi }\right) }+i(\mathcal{F}_{AB}^{(k)}+%
\mathcal{F}_{2}^{(k)})\hat{B}e^{it\tilde{\delta}_{+}}\right] ~
\end{equation}%
\begin{equation}
\hat{B}_{h}=\exp [i\mathcal{F}_{B}^{(k)}]\times \left[ \hat{B}e^{it\tilde{%
\delta}_{+}}+i(\mathcal{F}_{AB}^{(k)\ast }+\mathcal{F}_{2}^{(k)\ast })\hat{A}%
e^{it\left( \tilde{\delta}_{+}+\delta _{\chi }\right) }\right] ~,
\end{equation}%
where small time-dependent c-number functions are:
\begin{equation}
\mathcal{F}_{A}^{(\Omega )}=\varepsilon _{\Omega }\frac{\tilde{g}_{0}^{2}}{%
2\beta ^{2}}\left[ 2\frac{e^{it\eta }-1}{\eta }-\frac{e^{it\left( \eta
+\beta \right) }-1}{\eta +\beta }-\frac{e^{it\left( \eta -\beta \right) }-1}{%
\eta -\beta }+c.c.\right]
\end{equation}%
\begin{equation}
\mathcal{F}_{A}^{(g)}=\tilde{\varepsilon}_{g}\frac{\tilde{g}_{0}\Delta _{-}}{%
2\beta ^{2}}\left[ 2\frac{e^{it\eta }-1}{\eta }-\frac{e^{it\left( \eta
+\beta \right) }-1}{\eta +\beta }-\frac{e^{it\left( \eta -\beta \right) }-1}{%
\eta -\beta }+c.c.\right]
\end{equation}%
\begin{equation}
\mathcal{F}_{B}^{(\Omega )}=\varepsilon _{\Omega }\frac{\tilde{g}_{0}^{2}}{%
2\beta ^{2}}\left[ \left( 2+\frac{\Delta _{-}^{2}}{\tilde{g}_{0}^{2}}\right)
\frac{e^{it\eta }-1}{\eta }+\frac{e^{it\left( \eta +\beta \right) }-1}{\eta
+\beta }+\frac{e^{it\left( \eta -\beta \right) }-1}{\eta -\beta }+c.c.\right]
\end{equation}%
\begin{equation}
\mathcal{F}_{B}^{(g)}=\tilde{\varepsilon}_{g}\frac{\tilde{g}_{0}\Delta _{-}}{%
2\beta ^{2}}\left[ -2\frac{e^{it\eta }-1}{\eta }+\frac{e^{it\left( \eta
+\beta \right) }-1}{\eta +\beta }+\frac{e^{it\left( \eta -\beta \right) }-1}{%
\eta -\beta }+c.c.\right]
\end{equation}%
\begin{equation}
\mathcal{F}_{AB}^{(\Omega )}=\varepsilon _{\Omega }\frac{\tilde{g}_{0}}{%
2\beta ^{2}}\sum_{{\mathcal{S}}=\pm }\left[ -\Delta _{-}\frac{e^{{\mathcal{S}%
}it\eta }-1}{\eta }+\beta _{+}\frac{e^{{\mathcal{S}}it\left( \eta +{\mathcal{%
S}}\beta \right) }-1}{\eta +{\mathcal{S}}\beta }-\beta _{-}\frac{e^{{%
\mathcal{S}}it\left( \eta -{\mathcal{S}}\beta \right) }-1}{\eta -{\mathcal{S}%
}\beta }\right]
\end{equation}%
\begin{equation}
\mathcal{F}_{AB}^{(g)}=\tilde{\varepsilon}_{g}\frac{\Delta _{-}}{2\beta ^{2}}%
\sum_{{\mathcal{S}}=\pm }\left[ 4\delta _{-}\frac{e^{{\mathcal{S}}it\eta }-1%
}{\eta }+\beta _{+}\frac{e^{{\mathcal{S}}it\left( \eta +{\mathcal{S}}\beta
\right) }-1}{\eta +{\mathcal{S}}\beta }-\beta _{-}\frac{e^{{\mathcal{S}}%
it\left( \eta -{\mathcal{S}}\beta \right) }-1}{\eta -{\mathcal{S}}\beta }%
\right]
\end{equation}%
\begin{eqnarray}
\mathcal{F}_{2}^{(\Omega )} &=&\varepsilon _{\Omega }\frac{\tilde{g}_{0}}{%
2\beta ^{2}}\int_{0}^{t}d\tau \lbrack \mathcal{F}_{A}^{(\Omega )}(\tau )-%
\mathcal{F}_{B}^{(\Omega )}(\tau )] \\
&&\times \left( \sum_{{\mathcal{S}}=\pm }[-\Delta _{-}e^{{\mathcal{S}}i\tau
\eta }+\beta _{+}e^{{\mathcal{S}}i\tau \left( \eta +{\mathcal{S}}\beta
\right) }-\beta _{-}e^{{\mathcal{S}}i\tau \left( \eta -{\mathcal{S}}\beta
\right) }]-c.c.\right)  \nonumber
\end{eqnarray}%
\begin{eqnarray}
\mathcal{F}_{2}^{(g)} &=&\tilde{\varepsilon}_{g}\frac{\Delta _{-}}{2\beta
^{2}}\int_{0}^{t}d\tau \lbrack \mathcal{F}_{A}^{(g)}(\tau )-\mathcal{F}%
_{B}^{(g)}(\tau )] \\
&&\times \left( \sum_{{\mathcal{S}}=\pm }[4\delta _{-}e^{{\mathcal{S}}i\tau
\eta }+\beta _{+}e^{{\mathcal{S}}i\tau \left( \eta +{\mathcal{S}}\beta
\right) }-\beta _{-}e^{{\mathcal{S}}i\tau \left( \eta -{\mathcal{S}}\beta
\right) }]-c.c.\right) .  \nonumber
\end{eqnarray}%
One can check that to the first order in $\varepsilon _{\Omega }$ and $%
\tilde{\varepsilon}_{g}$ the slowly-varying operators $\hat{A}$ and $\hat{B}$
also satisfy the bosonic commutation relations.

Under resonant modulations the new operators $\hat{A}$ and $\hat{B}$ evolve
according to the effective Hamiltonians $\hat{H}_{1}^{(k)}$ and $\hat{H}%
_{2}^{(k)}$ for the 1- and 2-order resonances, respectively. The approximate
expressions for $\hat{H}_{1,2}^{(k)}$ are given below in the resonant and
dispersive regimes. For the sake of clarity we omit the non-Gaussian
(quartic) contributions arising from $\hat{H}_{NG}$ in equation (\ref{NG}),
as such terms were studied thoroughly in \cite{igor} for the 1-order
resonances.

\subsection{Resonant regime}

In the regime $\Delta _{-}=0$ we have (for the sake of space, in all the
expressions below we omit the \emph{hermitian conjugate} on the right-hand
side of the expressions):

\begin{itemize}
\item for $\eta \simeq 2\omega _{0}$ or $\eta \simeq \omega _{0}$%
\begin{equation}
\hat{H}_{1}^{(\Omega )}\simeq i\tilde{g}_{0}\frac{\tilde{g}_{0}}{2\omega
_{0}}\left( \frac{\varepsilon _{\Omega }}{4\Omega _{0}}\right) \left[ \hat{A}%
^{2}e^{2it\delta _{\chi }}-\hat{B}^{2}\right] e^{-it(2\omega _{0}-2\tilde{%
\delta}_{+}-\eta )}
\end{equation}%
\begin{equation}
\hat{H}_{2}^{(\Omega )}\simeq -\tilde{g}_{0}\frac{\tilde{g}_{0}}{2\omega
_{0}}\left( \frac{\varepsilon _{\Omega }}{4\Omega _{0}}\right) ^{2}8\left[
\hat{A}^{2}e^{2it\delta _{\chi }}-\hat{B}^{2}\right] e^{-it2(\omega _{0}-%
\tilde{\delta}_{+}-\eta )}
\end{equation}%
\begin{equation}
\hat{H}_{1}^{(g)}\simeq 0~,~\hat{H}_{2}^{(g)}\simeq 0\,;
\end{equation}

\item for $\eta \simeq 2(\omega _{0}\pm \tilde{g}_{0})$ or $\eta \simeq
\omega _{0}\pm \tilde{g}_{0}$%
\begin{equation}
\hat{H}_{1}^{(\Omega )}\simeq i\tilde{g}_{0}\left( \frac{\varepsilon
_{\Omega }}{8\Omega _{0}}\right) \left[ \hat{A}\hat{B}e^{it\delta _{\chi
}}\pm \frac{1}{2}\hat{A}^{2}e^{2it\delta _{\chi }}\pm \frac{1}{2}\hat{B}^{2}%
\right] e^{-it(2\omega _{0}\pm 2\tilde{g}_{0}-2\tilde{\delta}_{+}-\eta )}
\end{equation}%
\begin{equation}
\hat{H}_{2}^{(\Omega )}\simeq -\tilde{g}_{0}\left( \frac{\varepsilon
_{\Omega }}{8\Omega _{0}}\right) ^{2}\frac{1}{4}\left[ 14\hat{A}\hat{B}%
e^{it\delta _{\chi }}\pm 5\hat{A}^{2}e^{2it\delta _{\chi }}\pm 5\hat{B}^{2}%
\right] e^{-it2(\omega _{0}\pm \tilde{g}_{0}-\tilde{\delta}_{+}-\eta )}
\end{equation}%
\begin{equation}
\hat{H}_{1}^{(g)}\simeq -i\tilde{g}_{0}\frac{1}{2}\left( \frac{\tilde{%
\varepsilon}_{g}}{2\tilde{g}_{0}}\right) \left[ \hat{A}\hat{B}e^{it\delta
_{\chi }}\pm \frac{1}{2}\hat{A}^{2}e^{2it\delta _{\chi }}\pm \frac{1}{2}\hat{%
B}^{2}\right] e^{-it(2\omega _{0}\pm 2\tilde{g}_{0}-2\tilde{\delta}_{+}-\eta
)}
\end{equation}%
\begin{equation}
\hat{H}_{2}^{(g)}\simeq \tilde{g}_{0}\frac{1}{2}\left( \frac{\tilde{%
\varepsilon}_{g}}{2\tilde{g}_{0}}\right) ^{2}\frac{2\tilde{g}_{0}}{\omega
_{0}}\left[ \pm \hat{A}\hat{B}e^{it\delta _{\chi }}+\frac{1}{2}\hat{A}%
^{2}e^{2it\delta _{\chi }}+\frac{1}{2}\hat{B}^{2}\right] e^{-it2(\omega
_{0}\pm \tilde{g}_{0}-\tilde{\delta}_{+}-\eta )}\,.
\end{equation}
\end{itemize}

\subsection{Dispersive regime}

In the regime $\left\vert \Delta _{-}\right\vert /2\gg \left\vert \tilde{g}%
_{0}\right\vert $ we have

\begin{itemize}
\item for $\eta \simeq \Delta _{+}$ or $\eta \simeq \Delta _{+}/2$ ({\em AJC-like
behavior})%
\begin{equation}
\hat{H}_{1}^{(\Omega )}\simeq i\tilde{g}_{0}\left( \frac{\varepsilon
_{\Omega }}{2\Delta _{+}}\right) \left[ \hat{A}\hat{B}e^{it\delta _{\chi }}-%
\frac{\tilde{g}_{0}}{\Delta _{-}}\hat{A}^{2}e^{2it\delta _{\chi }}+\frac{%
\tilde{g}_{0}}{\Delta _{-}}\hat{B}^{2}\right] e^{-it(\Delta _{+}-2\tilde{%
\delta}_{+}-\eta )}
\end{equation}%
\begin{equation}
\hat{H}_{2}^{(\Omega )}\simeq -\tilde{g}_{0}\left( \frac{\varepsilon
_{\Omega }}{2\Delta _{+}}\right) ^{2}2\left[ \hat{A}\hat{B}e^{it\delta
_{\chi }}-\frac{\tilde{g}_{0}}{\Delta _{-}}\hat{A}^{2}e^{2it\delta _{\chi }}+%
\frac{\tilde{g}_{0}}{\Delta _{-}}\hat{B}^{2}\right] e^{-it(\Delta _{+}-2%
\tilde{\delta}_{+}-2\eta )}
\end{equation}%
\begin{equation}
\hat{H}_{1}^{(g)}\simeq -i\tilde{g}_{0}\left( \frac{\tilde{\varepsilon}%
_{g}}{2\tilde{g}_{0}}\right) \left[ \hat{A}\hat{B}e^{it\delta _{\chi }}-%
\frac{\tilde{g}_{0}}{\Delta _{-}}\hat{A}^{2}e^{2it\delta _{\chi }}+\frac{%
\tilde{g}_{0}}{\Delta _{-}}\hat{B}^{2}\right] e^{-it(\Delta _{+}-2\tilde{%
\delta}_{+}-\eta )}
\end{equation}%
\begin{equation}
\hat{H}_{2}^{(g)}\simeq -\tilde{g}_{0}\left( \frac{\tilde{\varepsilon}_{g}%
}{2\tilde{g}_{0}}\right) ^{2}\left( \frac{2\tilde{g}_{0}}{\Delta _{+}}%
\right) ^{2}5\left[ \hat{A}\hat{B}e^{it\delta _{\chi }}-\frac{\tilde{g}_{0}}{%
\Delta _{-}}\hat{A}^{2}e^{2it\delta _{\chi }}+\frac{\tilde{g}_{0}}{\Delta
_{-}}\hat{B}^{2}\right] e^{-it(\Delta _{+}-2\tilde{\delta}_{+}-2\eta )};
\end{equation}

\item for $\eta \simeq 2\omega _{0}$ or $\eta \simeq \omega _{0}$ ({\em DCE
behavior})%
\begin{equation}
\hat{H}_{1}^{(\Omega )}\simeq i\tilde{\delta}_{-}\frac{\Omega _{0}}{%
\Delta _{+}}\left( \frac{\varepsilon _{\Omega }}{2\Omega _{0}}\right) \left[
\hat{A}^{2}e^{2it\delta _{\chi }}+\frac{2\tilde{g}_{0}}{\Delta _{-}}\hat{A}%
\hat{B}e^{it\delta _{\chi }}+\frac{\tilde{g}_{0}^{2}}{\Delta _{-}^{2}}\hat{B}%
^{2}\right] e^{-it(2\omega _{0}+2\tilde{\delta}_{-}-2\tilde{\delta}_{+}-\eta
)}
\end{equation}%
\begin{equation}
\hat{H}_{2}^{(\Omega )}\simeq -\tilde{\delta}_{-}\frac{\Omega _{0}}{%
\Delta _{+}}\left( \frac{\varepsilon _{\Omega }}{2\Omega _{0}}\right) ^{2}%
\left[ \hat{A}^{2}e^{2it\delta _{\chi }}+\frac{4\tilde{g}_{0}}{\Delta _{-}}%
\hat{A}\hat{B}e^{it\delta _{\chi }}+\frac{3\tilde{g}_{0}^{2}}{\Delta _{-}^{2}%
}\hat{B}^{2}\right] e^{-it2(\omega _{0}+\tilde{\delta}_{-}-\tilde{\delta}%
_{+}-\eta )}
\end{equation}%
\begin{equation}
\hat{H}_{1}^{(g)}\simeq -i\frac{2\tilde{\delta}_{-}\Omega _{0}}{\Delta
_{+}}\left( \frac{\tilde{\varepsilon}_{g}}{2\tilde{g}_{0}}\right) \left[
\hat{A}^{2}e^{2it\delta _{\chi }}+\frac{2\tilde{g}_{0}}{\Delta _{-}}\hat{A}%
\hat{B}e^{it\delta _{\chi }}+\frac{\tilde{g}_{0}^{2}}{\Delta _{-}^{2}}\hat{B}%
^{2}\right] e^{-it(2\omega _{0}+2\tilde{\delta}_{-}-2\tilde{\delta}_{+}-\eta
)}
\end{equation}%
\begin{equation}
\hat{H}_{2}^{(g)}\simeq \frac{2\tilde{\delta}_{-}\Omega _{0}}{\Delta _{+}}%
\left( \frac{\tilde{\varepsilon}_{g}}{2\tilde{g}_{0}}\right) ^{2}\frac{%
\Delta _{+}\Delta_-}{2\Omega _{0}^2}\left[ \hat{A}%
^{2}e^{2it\delta _{\chi }}+\frac{2\tilde{g}_{0}}{\Delta _{-}}\hat{A}\hat{B}%
e^{it\delta _{\chi }}+\frac{\tilde{g}_{0}^{2}}{\Delta _{-}^{2}}\hat{B}^{2}%
\right] e^{-it2(\omega _{0}+\tilde{\delta}_{-}-\tilde{\delta}_{+}-\eta )}\,;
\end{equation}

\item for $\eta \simeq 2\Omega _{0}$ or $\eta \simeq \Omega _{0}$ we have an
effect analogous to DCE, but for the collective atomic excitations (so it
was called {\em Inverse-DCE} in \cite{igor})%
\begin{equation}
\hat{H}_{1}^{(\Omega )}\simeq -i\tilde{\delta}_{-}\frac{\omega _{0}}{%
\Delta _{+}}\left( \frac{\varepsilon _{\Omega }}{2\Omega _{0}}\right) \left[
\hat{B}^{2}-\frac{2\tilde{g}_{0}}{\Delta _{-}}\hat{A}\hat{B}e^{it\delta
_{\chi }}+\frac{\tilde{g}_{0}^{2}}{\Delta _{-}^{2}}\hat{A}^{2}e^{2it\delta
_{\chi }}\right] e^{-it(2\Omega _{0}-2\tilde{\delta}_{-}-2\tilde{\delta}%
_{+}-\eta )}
\end{equation}%
\begin{equation}
 H_{2}^{(\Omega )}\simeq \tilde{\delta}_{-}\frac{\omega _{0}}{\Delta _{+}}%
\left( \frac{\varepsilon _{\Omega }}{2\Omega _{0}}\right) ^{2}\left[ \hat{B}%
^{2}-\frac{4\tilde{g}_{0}}{\Delta _{-}}\hat{A}\hat{B}e^{it\delta _{\chi }}+%
\frac{3\tilde{g}_{0}^{2}}{\Delta _{-}^{2}}\hat{A}^{2}e^{2it\delta _{\chi }}%
\right] e^{-it2(\Omega _{0}-\tilde{\delta}_{-}-\tilde{\delta}_{+}-\eta )}
\end{equation}%
\begin{equation}
\hat{H}_{1}^{(g)}\simeq i\frac{2\tilde{\delta}_{-}\omega _{0}}{\Delta _{+}%
}\left( \frac{\tilde{\varepsilon}_{g}}{2\tilde{g}_{0}}\right) \left[ \hat{B}%
^{2}-\frac{2\tilde{g}_{0}}{\Delta _{-}}\hat{A}\hat{B}e^{it\delta _{\chi }}+%
\frac{\tilde{g}_{0}^{2}}{\Delta _{-}^{2}}\hat{A}^{2}e^{2it\delta _{\chi }}%
\right] e^{-it(2\Omega _{0}-2\tilde{\delta}_{-}-2\tilde{\delta}_{+}-\eta )}
\end{equation}%
\begin{equation}
\hat{H}_{2}^{(g)}\simeq \frac{2\tilde{\delta}_{-}\omega _{0}}{\Delta _{+}}%
\left( \frac{\tilde{\varepsilon}_{g}}{2\tilde{g}_{0}}\right) ^{2}\frac{%
\Delta _{+}\Delta_-}{2\omega _{0}^2}\left[ \hat{B}^{2}-%
\frac{2\tilde{g}_{0}}{\Delta _{-}}\hat{A}\hat{B}e^{it\delta _{\chi }}+\frac{%
\tilde{g}_{0}^{2}}{\Delta _{-}^{2}}\hat{A}^{2}e^{2it\delta _{\chi }}\right]
e^{-it2(\Omega _{0}-\tilde{\delta}_{-}-\tilde{\delta}_{+}-\eta )}\,.
\end{equation}
\end{itemize}

One can see that under the second-order resonances the $g$-modulation always
yields transition rates at least one order of magnitude smaller than the $%
\Omega $-modulation. Lastly, Anti-DCE for light and matter excitations
(or {\em Anti-Inverse-DCE}) can be derived from the Non-Gaussian Hamiltonian $\hat{H}%
_{NG}$ \cite{igor,lucas}, however we do not present the results because
under the 2-order resonances the corresponding transition rates are very
small (see section \ref{dr} for the case $N=1$).

\end{document}